\documentclass[lettersize,journal]{IEEEtran}
\usepackage{amsmath,amsfonts}
\usepackage{algorithmic}
\usepackage{algorithm}
\usepackage{array}
\usepackage[caption=false,font=normalsize,labelfont=sf,textfont=sf]{subfig}
\usepackage{hyperref,color,breqn,multirow}
\usepackage{textcomp,gensymb}
\usepackage{stfloats}
\usepackage{multirow}
\usepackage{booktabs}
\usepackage{url}
\usepackage{verbatim}
\usepackage{graphicx}
\usepackage{cite}
\usepackage{amssymb}
\makeatletter
\renewcommand*{\eqref}[1]{\hyperref[{#1}]{\textup{(\ref*{#1})}}}
\newcommand*{\figref}[1]{\hyperref[{#1}]{\textup{Fig.~\ref*{#1}}}}
\newcommand*{\tabref}[1]{\hyperref[{#1}]{\textup{Table~\ref*{#1}}}}
\newcommand*{\secref}[1]{\hyperref[{#1}]{\textup{Section~\ref*{#1}}}}
\newcommand*{\mat}[1]{\overline{\overline#1}}
\makeatother
\begin{document}


\title{All-Angle Scanning Leaky-Wave Antennas and Surface-Wave Routing by Reconfigurable Metasurfaces}

\author{Yongming Li, Xikui Ma, Viktar Asadchy,~\IEEEmembership{Senior Member,~IEEE}, and Sergei A.  Tretyakov,~\IEEEmembership{Fellow,~IEEE}

\thanks{This work was in part supported by the China Scholarship Council under Grant 202106280229, and in part by the Research Council of Finland under Project 345178. \textit{(Corresponding author: Yongming Li)}}

\thanks{Yongming Li is with the State Key Laboratory of Electrical Insulation and
Power Equipment, School of Electrical Engineering, 
Xi'an Jiaotong University, Xi'an 710049, China, and with the Department of Electronics
and Nanoengineering, Aalto University, P.O. Box 15500, FI-00076, Aalto, Finland (e-mail: yongmingli@stu.xjtu.edu.cn).}

\thanks{Xikui Ma is with the State Key Laboratory of Electrical Insulation and
Power Equipment, School of Electrical Engineering, 
Xi'an Jiaotong University, Xi'an, Shaanxi 710049, China.}

\thanks{Viktar Asadchy and Sergei A. Tretyakov are with the Department of Electronics
and Nanoengineering, Aalto University, P.O. Box 15500, FI-00076, Aalto, Finland.}

}



\maketitle

\begin{abstract}
In this work, 
we show that propagating waves can be fully converted into surface waves and back using geometrically periodic arrays of simple electrically small metal elements loaded by adjustable reactive loads. The proposed approach allows creation of all-angle scanning leaky-wave antennas with perfect or even superdirective aperture efficiency at all scan angles. Moreover, it is possible to co-design such leaky-wave antenna arrays with surface-wave waveguides that can guide the received power to the load or to another leaky-wave antenna section. That second section can either reradiate the received power into any direction or perform some other transformation of the reradiated wave front, for example, focusing the power at a point.  These and other  functionalities are realized by global optimization of reactive loads of array elements. This global optimization together with the use of arrays with a subwavelength geometrical period allows proper control over both propagating and evanescent-field distributions, ensuring theoretically perfect performance at arbitrary scan angles. The proposed technique can be used in antenna engineering and in advanced designs of reconfigurable intelligent surfaces.  

\end{abstract}


\section{Introduction}
\IEEEPARstart{I}{n} the domain of electromagnetic wave manipulation, metasurfaces play an invaluable role, including versatile control of propagating~\cite{wan2014simultaneous, ra2018reconfigurable, Kwon2018lossless, zhu2023bifunctional, liu2023reflectarrays} and evanescent waves~\cite{wan2014simultaneous, diaz2017generalized, wang2018extreme, Tcvetkova2018near, Popov2019omega}. One important functionality is the conversion between propagating and surface (guided) waves~\cite{Kwon2018arbitrary, Tcvetkova2018near}. Due to the momentum mismatch between propagating waves and surface waves, the conventional realization methods, such as prism~\cite{otto1968excitation} and grating~\cite{neviere1973about} coupling methods, suffer from inherently low conversion efficiency~\cite{sun2016high}. At  microwave frequencies, 
this conversion is usually realized by leaky-wave antennas (LWAs) ~\cite{jackson2008leaky,Tcvetkova2018near,Tcvetkova2019exact,Tcvetkova2020perfect,abdo2019leaky,lee2021large,bodehou_direct_2022,budhu_design_2022}. Typically, the radiated field at the aperture of an LWA does not exhibit a homogeneous in space amplitude nor a linearly varying phase profile. Hence, the conversion between the guided and propagating waves is not perfect in this scenario. Only recently aperiodic non-local LWAs were proposed that rigorously account for the mutual coupling between the radiating elements, providing the opportunity for near-perfect conversion efficiencies~\cite{xu_wideangle_2022}.

Conversion of a propagating wave to a surface wave was also reported based on \textit{periodic} metasurfaces~\cite{sun2012gradient, Mohammadi2016Wave,sun2016high}. The designed phase gradient of the metasurface partially compensated for the momentum mismatch at the interface between the two waves, however, the operating efficiency was very low.  In Ref.~\cite{Tcvetkova2018near}, Tcvetkova \textit{et al.} proposed an approximate solution for the periodic surface impedance  
capable of providing a nearly ideal conversion of a plane wave into a surface wave. In that work, a surface wave with quasi-linearly growing power is realized by using an anisotropic metasurface. However, the required impedance profile of such a design is hard to realize and tune due to its anisotropic nature. To overcome this problem, in Ref.~\cite{Popov2019omega}, Popov \textit{et al.} proposed an alternative method by using an omega-bianisotropic metasurface to realize such a metasurface converter without exploiting anisotropy. 
However, in that work, to achieve high conversion efficiency from propagating to surface wave, an input surface wave had to be excited at the same time. 
Furthermore, for periodic structures, the conversion efficiency is influenced by both the period and the incident angles. Any change in the incident angle or the period necessitates redesigning the structure to maintain high conversion efficiency. 

Once the surface wave is generated, it is crucial to guide it and at some point relaunch it back into free space~\cite{wan2014simultaneous,achouri2016surfacewave,achouri2018space}.
In Ref.~\cite{wan2014simultaneous},  the authors proposed the hybrid metasurface concept to control the surface wave and propagating wave simultaneously, where a planar metamaterial is used to interact with surface waves and a holographic metasurface is used to modulate propagating waves. 
In Refs.~\cite{achouri2016surfacewave,achouri2018space}, the authors demonstrated the conversion of plane into surface waves, their routing, and converting them back into an outgoing plane wave. However, their design suffered from low efficiency (about $10\%$) due
to the surface-wave dissipation loss, scattering at the metasurface
discontinuities, and limited conversion efficiency. 
In this work, we propose \textit{aperiodic} finite-sized metasurfaces that can convert between plane and surface waves, guide the surface wave, and relaunch it as a plane wave in a desired direction or focus it at a specific point, all with nearly 100\% efficiency. Beam scanning and full reconfigurability are achieved within a geometrically fixed platform, but properly adjusting variable bulk loads of the array elements. The metasurface is modeled as a realistic array of strips with loaded impedances which can be implemented using an established design procedure~\cite{wang2018systematic}. The values of the load impedances are determined using the analytical theory empowered by the multi-population genetic algorithm (MPGA) optimization~\cite{katoch2021review,slowik2020evolutionary}.
The proposed metasurfaces are modular, meaning they can be designed individually and cascaded in any order without the need for global optimization.

Interestingly, a similar concept of plane-to-surface wave conversion was recently proposed in Refs.~\cite{kwon_arbitrary_2018,kwon_modulated_2020,budhu2024near,arshed2024direct}. However, the approach presented in \cite{kwon_arbitrary_2018,kwon_modulated_2020,budhu2024near} relies on the ideal sheet impedance description and requires subsequent global optimization. The resulting optimized impedance profile can exhibit sharp spatial variations and include impedance values that are difficult to realize in practice. Moreover, this method is less computationally efficient and fails to account for the coupling between elements in the metasurface when implemented with realistic patterned unit cells~\cite{budhu2023unit}. The approach proposed in~\cite{arshed2024direct}, while enabling nearly 100\% conversion efficiencies, is limited to electrically large metasurfaces with slowly varying modulation of their surface properties, as it relies on the locally periodic problem. 
In sharp contrast, our approach rigorously accounts for the coupling between the unit cells using the mutual impedance matrix, and global optimization is not required. Moreover, since the metasurface is composed of discrete elements (strips), the approach can also support rapidly varying impedance profiles, enabling the design of realistic and compact metasurface converters. Finally,  after being fabricated, the proposed metasurfaces, empowered by tunable reactive loads such as varactor diodes, can maintain perfect conversion efficiency, even when the angle of incidence (or the angle of radiation in the leaky-mode regime) varies \textit{continuously}.

The remaining parts of this work are organized as follows. In~\secref{sec:section2}, we present the theoretical framework. In~\secref{sec:section3}, efficient conversion from a plane wave to surface waves is obtained followed by the optimization of the surface wave spectrum in~\secref{sec:section4}. The reciprocal operation of the metasurfaces is analyzed in~\secref{sec:section5}. In~\secref{sec:section6}, the conversion from a Gaussian beam to a surface wave is presented. 
Next, in~\secref{sec:section7}, we present examples demonstrating the guiding of generated surface waves and their reradiation into free space, including possibilities for beam scanning and near-field focusing. Concluding remarks are provided in~\secref{sec:section8}.

\section{Principle and methodology}
\label{sec:section2}
\begin{figure}
    \centering    
\includegraphics[width=0.95\columnwidth]{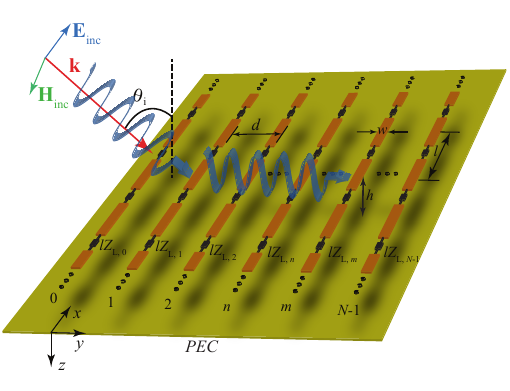}
    \caption{(a) A finite-sized array is composed of $N$ metallic loaded strips over an infinite ground plane under a TE-polarized plane-wave illumination at an angle $\theta_{\rm i}$. The metasurface is designed to convert the incident plane wave into a surface wave. 
    The strips are loaded by reconfigurable bulk complex-valued impedances inserted periodically with the period $l$.}
    \label{fig:configuration}
\end{figure}

We model the aperiodic metasurface as a \textit{finite} array of thin conducting strips oriented along the $x$-direction and located at a distance $h$ above an \textit{infinite} perfect electric conductor (PEC) ground plane. The configuration of the proposed structure is shown in \figref{fig:configuration}. The ground plane is in the $z=0$ plane. The strips of width $w$ are uniformly spaced by a distance $d$ along the $y$-axis. To be more specific, the very first strip is numbered  0 and placed at position $y=0$, while the $n$-th strip is placed at position $y_n=nd$. For the analysis, thin conducting strips can be modeled as equivalent round wires with the effective radius $r_{\rm eff} = w/4$~\cite{tretyakov2003analytical}. 
The strips are periodically loaded, with a period $l$, by load impedances $l Z_{{\rm L},n}$, where $Z_{{\rm L},n}$ represent the load impedances per unit length of the strips (see \figref{fig:configuration}). The load impedance per unit length is uniform along the $x$-direction. 
It is assumed that $l$ and $w$ are much smaller than the wavelength in free space. Hence, the periodically loaded strips can be modeled as homogeneous thin impedance strips with the impedances per unit length  $Z_{{\rm L}, n}$, where $n \in \{0, 1, 2, \cdots, N-1\}$.
While the geometric arrangement of the strips is periodic along the $y$-axis, the variation in load impedances along this axis renders the metasurface aperiodic in the $y$-direction.

The metasurface is illuminated by a plane wave at an incidence angle $\theta_{\rm i}$. The incident wave is assumed to be a transverse-electric (TE)-polarized wave with the electric field given by 
\begin{equation}
    \mathbf{E}^{\rm inc}(y,z) = E_0 e^{-j k_0( \sin \theta_{\rm i} y + \cos\theta_{\rm i} z)} \hat{x},
\end{equation}
where  $E_0$ is the amplitude of the incident wave and $k_0$ represents its wavenumber in free space. Throughout the paper, the time dependence of the complex-valued fields is assumed to be in the form of $e^{j \omega t}$. The specular reflection of the incident wave from the ground plane has the electric field $\mathbf{E}^{\rm ref} = - E_0 e^{-j k_0 (\sin \theta_{\rm i} y-\cos \theta_{\rm i} z)} \hat{x}$. The superposition of the incident wave and the specular reflection of the incident wave from the ground plane is denoted as the external field, where the value at coordinate $(y, z)$ is $E^{\rm ext}_x (y,z) = -j 2 E_0 \sin (k_0 \cos \theta_{\rm i} z) e^{-j k_0 \sin \theta_{\rm i}y}$.
In what follows, without loss of generality, the frequency and amplitude of the incident wave are chosen as $f = 10$~GHz, and $E_0 = 1$~V/m, respectively. The distance $h$ above the ground plane is chosen $\lambda/6$ and the width of the strips is $w=\lambda/100$. 
 
To calculate the scattered electric field from the considered metasurface, we use the approach proposed in \cite{li2024all} based on  Ohm's law and the load impedance matrix. This approach assumes that the effective radius of the strips is negligible compared to both the
wavelength ($k_0 r_{\rm eff} \ll 1$) and the distance between the strips ($r_{\rm eff} \ll d$), so that we can replace the field generated by each strip with that of a line current~\cite{tretyakov2003analytical}. For each loaded strip, the relationship between induced line current and induced voltage on the loaded strip has to satisfy Ohm's law. The relationship between the induced line current and the external voltage can be bridged through the load impedance matrix $\mat{Z}$ as
\begin{equation}
    \mat{Z} \cdot \vec{I} = \vec{U},
    \label{eq:matrix_Ohm}
\end{equation}
where the vector of induced currents is $\vec{I} = \left[ I_0, I_1, \cdots, I_{N-1} \right]^{\rm T}$   and the external voltage vector is $\vec{U} =  \left[ E^{\rm ext}_x(y_0,-h), E^{\rm ext}_x(y_1,-h), \cdots, E^{\rm ext}_x(y_{N-1},-h) \right]^{\rm T}$. $\mat{Z} = \mat{Z}_{\rm s} + \mat{Z}_{\rm L} +\mat{Z}_{\rm m}$ is composed of the diagonal self-impedance matrix $\mat{Z}_{\rm s}={\rm diag}\left(Z_0, Z_1, \cdots, Z_n, \cdots, Z_{N-1} \right)$, the diagonal load impedance matrix $\mat{Z}_{\rm L} = {\rm diag}(Z_{{\rm L}, 0}, Z_{{\rm L}, 1},\cdots, Z_{{\rm L}, N-1})$, and the mutual impedance matrix $\mat{Z}_{\rm m}$ with the matrix elements given by $Z_{nm}  = \small{\frac{k_0 \eta_0 }{4}  \left[ H_0^{(2)} \left(k_0 d_{nm} \right)  -  H_0^{(2)}\left(k_0 \sqrt{d_{nm}^2+ 4 h^2 } \right) \right]
    }$.

For a given set of load impedances $\mat{Z}_{\rm L} = {\rm diag}(Z_{\rm L,0}, Z_{\rm L,1}, \cdots, Z_{{\rm L},N-1})$, the corresponding induced current distribution can be obtained by solving \eqref{eq:matrix_Ohm}. 
Once the induced current distribution is known, the electric field generated by the strip array in the upper half-space can be calculated as
\begin{align}
   \mathbf{E}^{\rm strips}=&- \frac{k_0 \eta_0 }{4}  \sum_{m=0}^{N-1} 
 I_m \left[ H_0^{(2)} \left(k_0 \sqrt{\left( y - y_m \right)^2+ \left( z+h \right)^2 } \right) \right.\notag\\
 &\left. -  H_0^{(2)}\left(k_0 \sqrt{\left( y - y_m \right)^2+ \left(z-h \right)^2 } \right) \right] \hat{x}. 
 \label{eq:electric_field_strips}
\end{align}
The scattered electric field is calculated by subtracting the incident electric field from the total electric field. For our finite-sized array, the scattered electric field $\mathbf{E}^{\rm sca}$ is the sum of the field generated by the strips $\mathbf{E}^{\rm strips}$ and the specularly reflected field $\mathbf{E}^{\rm ref}$:
 \begin{equation}
     \mathbf{E}^{\rm sca} = \mathbf{E}^{\rm strips} + \mathbf{E}^{\rm ref}.
 \label{eq:scat}
 \end{equation}

In this work, multi-population genetic algorithm (MPGA) is used to optimize the load impedances. Compared with our earlier work~\cite{li2024all}, where the optimization was performed using function \textit{fmincon} in MATLAB, MPGA provides a more robust and flexible approach for handling complex, multi-modal objective landscapes, which are common in load impedance optimization problems. Unlike gradient-based methods, MPGA does not rely on initial guesses and is less likely to get trapped in local minima, making it well-suited for non-convex problems~\cite{katoch2021review,slowik2020evolutionary}. Moreover, MPGA inherently supports parallelism and population diversity, which enhances its global search capabilities and convergence reliability.

\begin{figure}
    \centering
    \includegraphics[]{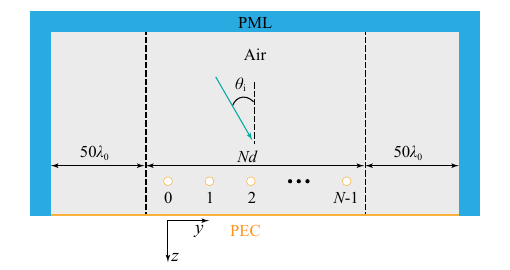}
    \caption{COMSOL Multiphysics Simulation configuration. The blue part is the PML (perfect matched layer), while the shaded area is the air background. The orange line at the bottom is PEC. The orange circles are used to simulate surface current density while connected with bulk loads under the illumination of incident wave with the incident angle $\theta_{\rm i}$.}
    \label{fig:comsol_configuration}
\end{figure}

To keep our design feasible, we introduce the following restriction on the imaginary part of the optimized impedances in all the strips: $-9 \times 10^5~\Omega/{\rm m} < \Im \{ Z_{{\rm L},n} \} < -500~\Omega/{\rm m}$. This restriction ensures that all the optimized strips will have a capacitive response with realistic capacitance values. 
Once the optimization is completed, we calculate the distribution of the scattered electric field and Poynting vector above the metasurface by simulating it in COMSOL Multiphysics. 
The simulation domain is shown in~\figref{fig:comsol_configuration}. The line currents with indices from 0 to $N-1$ are surrounded by perfectly matched layers on the three sides and a perfect electric conductor (PEC) boundary at the bottom. 

\section{Conversion from a plane wave to surface waves}
\label{sec:section3}

To obtain a high-efficiency conversion of the incident plane wave into a surface wave, in our optimization, we assume that all the strips of the metasurface except the last one are loaded with purely reactive loads, i.e., $Z_{{\rm L},n} = j X_{{\rm L},n}$ for $n \in [0; N-2]$, where $X_{{\rm L},n}$ denote reactance per unit length of the $n$-th strip. The last strip is described by a complex impedance $Z_{{\rm L},N-1} = R_{{\rm L},N-1}+ j X_{{\rm L},N-1}$, where $R_{{\rm L},N-1}$ is a positive-valued resistance per unit length, denoting dissipation loss. We optimize the values of the resistance $R_{{\rm L},N-1}$ and reactances $X_{{\rm L},n}$ with the objective to maximize the dissipated power in the last ($N-1$) strip. 
This objective will ensure that the power incident on all lossless strips will be directed along the metasurface toward its right edge and eventually terminated in the lossy load $Z_{{\rm L},N-1}$. Such channeling of the power along the metasurface automatically results in the efficient conversion of the incident plane wave into the surface (guided) waves before the whole incident power is absorbed in the last strip as in a perfectly matched transmission line. 
Importantly, in the following sections, we show that once optimized, this propagating-to-surface wave converting metasurface can be used as a module together with other metasurface modules, in particular, those that are designed to simply guide surface waves or reradiate them into free space. In this case, we can use the same optimized load impedance values and only eliminate the resistance $R_{{\rm L},N-1}$  of the last strip. 

Based on the aforementioned observations, we define the conversion efficiency of propagating into surface waves as the ratio of the power absorbed  in the last strip (collected by the perfectly matched load) and the incident power $P_{\rm inc}=\frac{E_0^2}{2 \eta} N d \cos \theta_{\rm i}$ illuminating the geometric area of the metasurface:
\begin{equation}
    \xi = \frac{P_{N-1}}{P_{\rm inc}}.
    \label{eq:conversion_efficiency}
\end{equation}
Here, $P_{N-1}=\frac{1}{2} \left| I_{N-1} \right|^2 \Re{ \{ Z_{{\rm L}, N-1}} \}$ represents the dissipated power by the last strip connected with lossy loads.  
Thus, to optimize the metasurface for the maximum conversion efficiency, we define the objective function as
\begin{equation}
    \mathcal{O} = {\rm min} \left\{ -P_{N-1} \right\}= {\rm min} \left\{  -\frac{1}{2} \left| I_{N-1} \right|^2 \Re \{ Z_{{\rm L},N-1}\} \right\}.
    \label{eq:objective_function}
\end{equation}

In this section, we fix the length of the metasurface to  $L = N d=6.5\lambda_0$ without loss of generality. 
First, we analyze the scenario where the incident plane wave is impinging on the metasurface at the normal direction, i.e. $\theta_{\rm i} = 0^\circ$. 
The distance between the adjacent strips in the metasurface $d$ strongly influences the achievable conversion efficiency. 
The optimized conversion efficiencies $\xi$ for different values of spacing $d$ are shown in~\tabref{tab:efficiency_M}. 
\begin{table}[!htbp]
    \caption{The plane-to-surface-wave conversion efficiency  of optimized metasurfaces with different distances between the strips}
    \centering
    \begin{tabular}[c]{ccccccc}
    \toprule
         $d/\lambda$   &  1/2&  1/4&  1/6&  1/8&  1/10&  1/12\\
    \midrule
         $\xi$ (\%) & 13.7&   79.6&   99.2&   101.5&  113.4 &  114.6 \\
    \bottomrule
    \end{tabular}
    \label{tab:efficiency_M}
\end{table}
When the distance between the adjacent strips is $d=\lambda/2$ (scenario of traditional reflectarrays), it is impossible to generate necessary evanescent waves (same as surface waves)~\cite{liu2023reflectarrays,li2024all}. In this case, the maximum conversion efficiency is $13.7\%$ (see \tabref{tab:efficiency_M}). To effectively excite the optimal spectrum of evanescent waves, the density of strips should be increased. It can be noted that  the conversion efficiency improves monotonically with the increase of the strip density, especially when the spacing between adjacent strips is reduced from $\lambda/2$ to $\lambda/4$, where evanescent waves are effectively controlled.
When the distance is $d=\lambda/8$ or less, the strip array exhibits the so-called superdirective conversion, that is, when the metasurface ``accumulates'' energy from a wider area of the incoming plane wavefront than its geometric area. A similar superdirective regime was recently demonstrated for finite-size electromagnetic absorbers of plane waves~\cite{li2024going,tretyakov2014maximizing, Maslovski2018Superabsorbing}. 
 
\begin{figure}
    \centering
    \includegraphics[width=0.95\columnwidth]{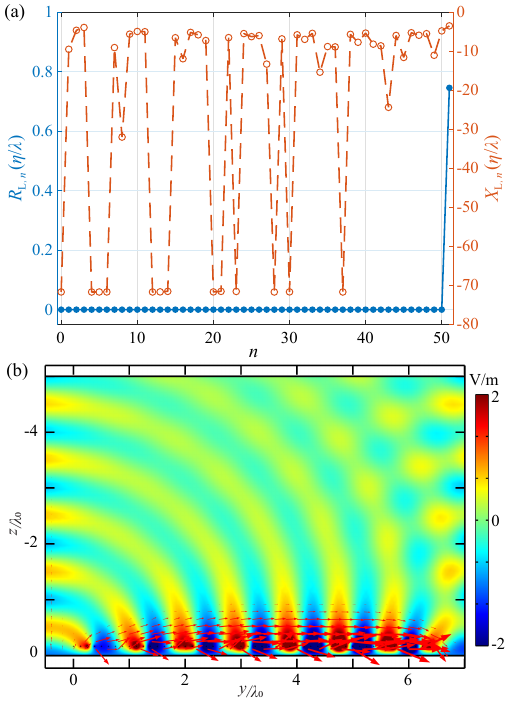}
    \caption{(a) Real and imaginary parts of the optimized load impedance as a function of strip number $n$ for the metasurface converting a normally incident wave ($\theta_{\rm i} = 0\degree$ and $|E^{\rm inc}| =1$~V/m) to surface waves. (b) Real part of the scattered electric field $\Re \{ E^{\rm sca} \}$~[V/m] above the metasurface. The red arrows indicate the Poynting vector distribution of the scattered wave. 
    }
    \label{fig:absorber_0deg}
\end{figure}

Although according to \tabref{tab:efficiency_M} further increasing the density of the strips in the metasurface leads to a saturating improvement in conversion efficiency, in what follows, we select $d=\lambda/8$ as the optimal value, balancing conversion efficiency with a reduced number of strips to facilitate simpler and more cost-effective fabrication. For this strip spacing, the conversion efficiency calculated by \eqref{eq:conversion_efficiency} reaches $101.5\%$.
 The optimized load impedances of such  metasurface are depicted in \figref{fig:absorber_0deg}(a).  As one can see, all the impedances have capacitive responses. Their capacitances do not reach extreme values, which makes them feasible for implementation. Although the variation of the impedance values between neighboring current wires is abrupt, it does not introduce any challenge for the implementation since our theory already takes into account the couplings between all the current strips. 
The scattered electric field by the optimized metasurface is plotted in \figref{fig:absorber_0deg}(b). This field includes the scattering by both the strips and the ground plane. 
The scattered field is confined near the surface of the strip array, which means that the incident plane wave is indeed effectively converted to a surface wave prior to being absorbed in the terminating strip of the array. The incident power illuminating the metasurface is efficiently channeled along the metasurface as a guided surface mode, as seen from the Poynting vector distribution shown by red arrows in \figref{fig:absorber_0deg}(b). 
Due to the accumulation of power, the converted surface wave is not uniform across the metasurface. The intensity of the electric field becomes stronger as the wave propagates along the surface and diminishes gradually near the lossy last strip of the metasurface. Almost all the  energy of the converted surface wave is eventually dissipated by the lossy loads connected to the last strip, which behaves like a perfectly matched load of a transmission line.  

\begin{figure}[!htbp]
    \centering    \includegraphics{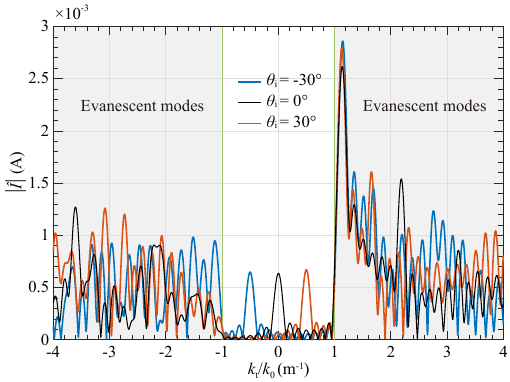}
    \caption{The complex amplitudes of the spatial harmonics of the electric current on the metasurface ($z=-h$) for three different incident angle scenarios.  The shaded area represents the evanescent modes  while the propagating modes exist in the white area (inside the light cone). The vertical green solid lines are the light lines. The positive values of $k_{\rm t}$ mean that the generated wave by the strip array propagates towards the $+y$-axis, while the negative one means propagating towards the $-y$-axis.}    \label{fig:diff_inc_angle_kspace}
\end{figure}

To illustrate the spectrum of wavenumbers of the generated surface waves (evanescent modes) at the metasurface, we calculate the spatial Fourier transform of the total induced currents along the surface of the strip array $I(y) = \sum_{n=0}^{N-1} I_n \delta(y-y_n)$  given by:
\begin{equation}
    \hat{I}(k_{\rm t}) = \int_{-\infty}^{+\infty} I(y) e^{-j k_{\rm t} y} dy = \sum_{n=0}^{N-1} I_n e^{-j k_{\rm t} y_n},
\end{equation}
where  $k_{\rm t}$ represents the tangential component of the wavevector. 
The amplitude of the induced current $\hat{I}$ as a function of $k_{\rm t}$ is shown in~\figref{fig:diff_inc_angle_kspace}. For normal incidence, it is represented by a black solid curve in the figure. The shaded region depicts the evanescent modes (modes outside the light cone). There is a single peak in the free-space-propagation region precisely at $k_{\rm t}=0$. This peak corresponds to induced space-uniform currents that generate a propagating reflected wave that destructively interferes with the wave specularly reflected by the ground plane, nearly completely eliminating it.  The peak outside the light cone, close to its right edge,  corresponds to the dominant surface mode with the normalized wavenumber $k_{\rm t}/k_0=1.135$. 
It is important to note that there are other smaller-amplitude peaks in the region of evanescent modes, implying that multiple surface waves are generated along the metasurface.  

We have also optimized the metasurfaces for converting obliquely incident waves  ($\theta_{\rm i}=  30\degree$ and $\theta_{\rm i}=  -30\degree$) into surface waves. 
For the former and latter scenarios, after the optimization, we obtained  the conversion efficiencies according to~\eqref{eq:conversion_efficiency} to be $115.4\%$ and $110.5\%$, respectively. These values are even higher than for the normal illumination scenario.

\section{Controlling the wavenumber spectrum of the surface modes}
\label{sec:section4}
As is seen in~\figref{fig:diff_inc_angle_kspace}, the generated surface wave is not a single mode but a group of surface modes, in which the forward and backward traveling surface wave co-exist. Indeed, although most of the power is carried by surface modes with wavenumber $k_t/k_0=1.135$, there is also strong (especially for $\theta_{\rm inc}=-30^\circ$) power carried in the opposite direction with $k_t/k_0= -1.135$. This superposition of the forward and backward surface modes leads to a partial standing wave pattern at the metasurface. 
It may be expected that the best conversion scenario would be creation of a single surface mode that would carry the power to the matched load. However, interestingly, the optimization shows that it is optimal to allow some reflections from the load. In this optimal regime, the array behaves as a resonant object, where the reactive fields interact with the incident field at longer times, leading to higher, even superdirective conversion efficiencies. It is important that in this case the resonant object is the whole array, potentially offering better bandwidth compared to more conventional superdirective arrays formed by many small resonant elements. 

However, the use of resonant metasurface panels may be not desirable in practical situations. Next, we show that this resonance is effectively suppressed by introducing small real parts to the load impedances inside the first $N-1$ strips. These non-zero real parts could be due to dissipative loss in the loads and/or due to the finite conductivity of metallic strips. In the latter scenario, the real part of the load impedance can be estimated by $\Re\{Z_{\rm L}\} =  \frac{\rho }{S_{\rm eff}}$, where $\rho$ represents the bulk resistivity of the strips, while the effective area $S_{\rm eff} = 2 \pi r_{\rm eff}\delta$ represents the effective skin surface area of the strip with the skip depth given by $\delta = \sqrt{\frac{2\rho}{\omega \mu_0}}$. For example, for copper strips with $r_{\rm eff}=w/4 =\lambda/400$, at the frequency of 10~GHz, the parasitic real parts of the impedance per unit length reaches the value of $\Re\{Z_{\rm L}\} =55 \Omega/{\rm m}$.

\begin{figure}[!ht]
    \centering\includegraphics{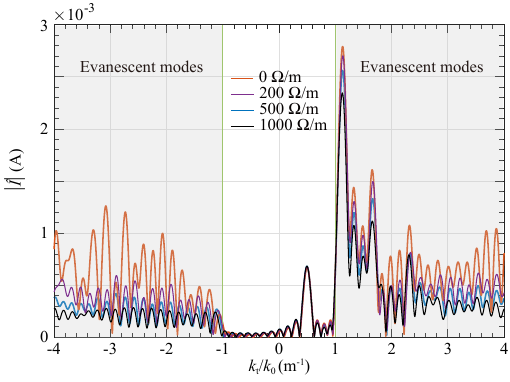}
    \caption{The complex amplitudes of the spatial harmonics of the electric current on the metasurface ($z=-h$)  for the introduced different lossy load impedance, when incident angle $\theta_{\rm inc}$ equal $30\degree$.  All the load impedances except the last one have the same real parts described in the plot legend.  }
    \label{fig:introduce_loss}
\end{figure}

In \figref{fig:introduce_loss}, we plot the spatial harmonics of the electric current for the optimized metasurface with $\theta_{\rm inc}=30 ^\circ$ when the resistance per unit length in the first $N-1$ strips is varied between $0 ~\Omega/ {\rm m}$ and $1000 ~\Omega/ {\rm m}$.
We can see that the backward propagating surface modes with $k_t/k_0= - 1.135$  are strongly suppressed even with very small resistance values. Expectedly, the introduction of the resistance leads to a small drop in the conversion efficiency of plane waves to surface waves (around 2\% for the scenario of 200~$\Omega/{\rm m}$).

\begin{figure}
    \centering
    \includegraphics{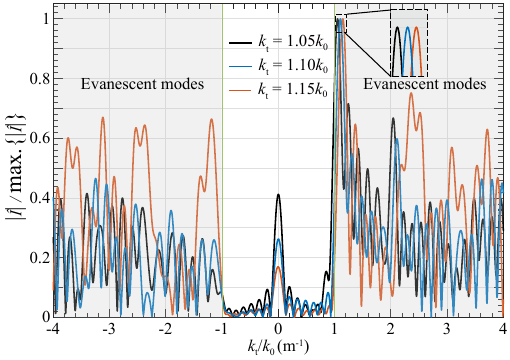}
    \caption{The complex amplitudes of the spatial harmonics of the electric current for metasurfaces designed to generate waves with different desired tangential wavevectors $k_{\rm des}$. The illumination is at normal incidence. }
    \label{fig:different_kt}
\end{figure}

Furthermore, it is possible to maximize the power carried by a specific mode with some  desired tangential component of the wavevector $k_{\rm des}$. This can be achieved using the objective function defined as
\begin{equation}
    \mathcal{O} = - f(k_{\rm des}) P_{N-1},
\end{equation}
where $\small{f(k) = {\rm ge}\left(\left| \hat{I} (k) \right|, \left| \hat{I} (k - \delta) \right|\right) {\rm ge} \left(\left| \hat{I} (k) \right| , \left| \hat{I} (k + \delta) \right| \right)}$ is used to control the value of $k_{\rm t}$,   $\delta$ is a small number chosen to be   $1.64$ in this study. Logical function ${\rm ge} (x_1,x_2)$ returns 1 if independent variables $x_1$ and $x_2$ satisfy $x_1 \ge x_2$, otherwise it returns 0. To show the effectiveness of the method, several examples with normal incidence are shown below. Function $f(k_{\rm des})$ returns 1 if local maximum points occurs at the position $k_{\rm des}$, otherwise it returns 0. To illustrate the effectiveness of this manipulation, the $k_{\rm des}$ is selected as $1.05k_0$, $1.10k_0$, and $1.15k_0$, respectively. The calculated magnitudes   of the induced current spatial harmonics  are depicted in~\figref{fig:different_kt}, respectively. The corresponding conversion efficiencies in these cases are 74.3\%, 94.1\%, and 97.8\%, respectively. As one can see, after the optimization, the desired current spatial harmonic was strongly enhanced. At the same time, the standing wave pattern became more prominent, especially for $k_{\rm des} =1.05k_0$.

\section{Metasurfaces operating as transmitting leaky-wave antennas}
\label{sec:section5}
When fed from the last strip, the designed metasurfaces are capable of launching a focused beam into free space due to the reciprocity principle~\cite{Asadchy2020tutorial}. Moreover, by reconfiguring the other strips, one can tune the propagation direction of this beam. 
Mechanical scanning methods of leaky-wave antennas require complex driven systems, suffering from high cost and slow scanning speed~\cite{jackson2008leaky}. Electronic scanning methods \cite{lw_steering2004,jiang_liquid_2019,raheemalsoad_electrically_2021} have limited beam scanning range and aperture efficiency. Here, two different electrically scanning methods are illustrated for the designed metasurfaces. One is realized by tuning the reconfigurable load impedances, keeping the working frequency fixed. Another method is realized by tuning the working frequency of the feeding network with fixed load impedances.


\subsection{Scanning method I: tuning the working frequency of the feeding network with fixed load impedances}
\begin{figure}
    \centering    \includegraphics{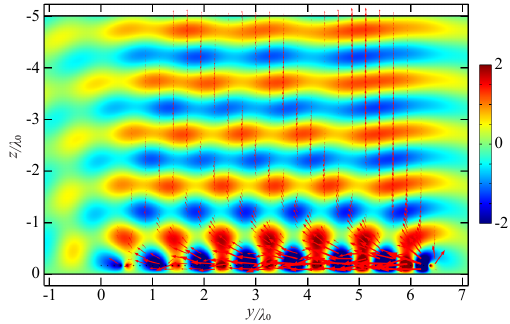}
    \caption{The real part of the scattered electric field $\Re\{E^{\rm sca}_x\}$~[V/m]  distributions over space. The red arrows represent the distribution of the Poynting vector. The operating frequency is 10~GHz.}    \label{fig:normal_incidence_sca_field_lwa}
\end{figure}

\begin{figure}
    \centering
    \includegraphics{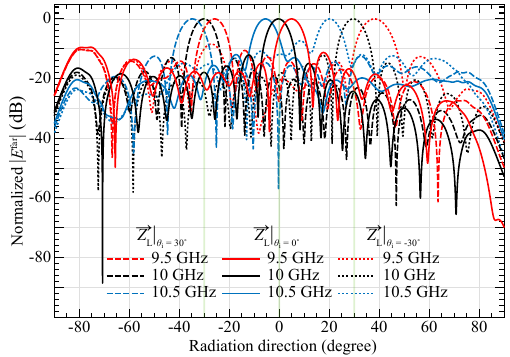}
    \caption{Radiation patterns for different working frequencies for three different metasurfaces. Initially, the metasurfaces were designed to convert incident plane waves at angles $\theta_{\rm i}=-30^\circ$, $0^\circ$, and $+30^\circ$ into a surface wave absorbed by the last strip. By reciprocity, in the transmitting regime, the metasurfaces were fed from the last strip, generating leaky waves with planar wavefronts propagating towards angles $\theta_{\rm rad}=-30^\circ$, $0^\circ$, and $+30^\circ$, respectively.
    }
    \label{fig:scanning_method1}
\end{figure}

Here the real parts of loads (resistors) are assumed to be independent of working frequency, while the imaginary parts (capacitors are assumed in this work) are dependent on the working frequency. For the load impedance  $Z_{{\rm L},n}$, it follows $Z_{{\rm L},n}= R_{{\rm L}, n} + 1/ (j\omega C_{{\rm L}, n})$. 
To create a leaky-wave antenna, we modify the optimized metasurface shown in Fig.~\ref{fig:absorber_0deg}(a) by equating the resistance of the last strip to zero, i.e., $R_{{\rm L}, N-1}=0 $, assuming an ideal voltage source feeding. Now, instead of illuminating the metasurface from the normal direction, we feed it from the last $(N-1)$-s strip. Due to reciprocity, the metasurface generates broadside radiation in the form of a beam leaking toward the normal direction. Although this beam has a quasi-planar wavefront, it is not a plane wave due to its finite waist (the metasurface is finite).  
The radiated electric field distribution above the metasurface is depicted in \figref{fig:normal_incidence_sca_field_lwa} at the working frequency of 10~GHz. The wavefront of the radiated wave indeed is nearly planar, exhibiting a space-uniform radiated beam by the entire metasurface area.
The radiation pattern of the metasurface is shown in \figref{fig:scanning_method1} with the black solid curve. Similarly to the conventional LWAs, the proposed structure also possesses main beam scanning capacity when the working frequency of the feeding network changes. As the working frequency increases from 9.5 GHZ to 10 GHz, the steered beam scans from the backward radiation across the broadside to the forward radiation (see \figref{fig:scanning_method1} with solid curves). This broadside radiation is realized due to the perfect impedance match, and the open stop band effects are prohibited.
 
\begin{figure}
    \centering
    \includegraphics{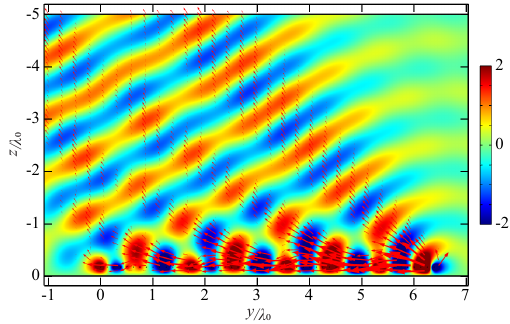}
    \caption{The real part of the scattered electric field $\Re\{E^{\rm sca}_x\}$~[V/m]  distributions over space. The red arrows represent the distribution of the Poynting vector.}
    \label{fig:30_sca_field_lwa}
\end{figure}

The metasurface can be also designed to operate as a  LWA in the 
forward-radiation and backward-radiation regimes~\cite{Jackson2008ch7,Monticone2015leaky}. The former regime is 
also referred to as an anomalous (or fast-wave) radiation regime. In this scenario,   the main radiation beam is directed away from the source (the last metasurface strip). The intensity of the scattered electric field and the Poynting vector show an exponential increase along the broadside direction. First, we optimize the metasurface to convert an incident plane wave from $\theta_{\rm i} = 30\degree$ into surface waves absorbed by the last strip. Next, we remove the resistance of the last strip and feed it with a wave port.  The radiated electric field and the Poynting vector distribution are shown in~\figref{fig:30_sca_field_lwa} for a working frequency of 10~GHz. One can observe the beam radiation  toward angle $\theta_{\rm rad}=30^\circ$. When scanning the frequency of the fed signal, the radiation direction can be conveniently tuned, as shown in \figref{fig:scanning_method1} with dashed curves. 

\begin{figure}
    \centering    
    \includegraphics{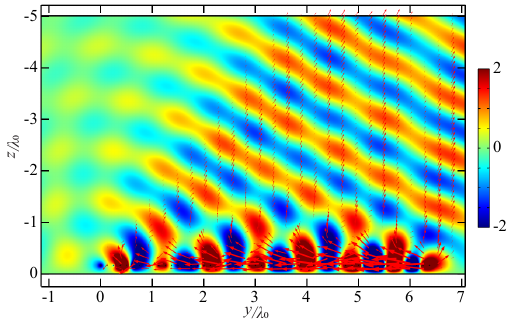}
    \caption{The real part of the scattered electric field $\Re\{E^{\rm sca}_x\}$~[V/m]  distributions over space. The red arrows represent the distribution of Poynting vectors.}
    \label{fig:negative30_sca_field_lwa}
\end{figure}

In the backward-radiation (or slow-wave) regime,  the main beam of radiation is directed backward, toward the source. For the metasurface designed to operate in this regime,   the radiated electric field and the Poynting vector are depicted in \figref{fig:negative30_sca_field_lwa}. 
The radiated electric field has an exponential decay along the $-z$-axis. 
The radiation pattern of the metasurface as the working frequency increases is represented by the dotted lines in~\figref{fig:scanning_method1}.



\subsection{Scanning method II: Tuning the reconfigurable load impedances at a fixed frequency}
\begin{figure}[!htbp]
    \centering
    \includegraphics{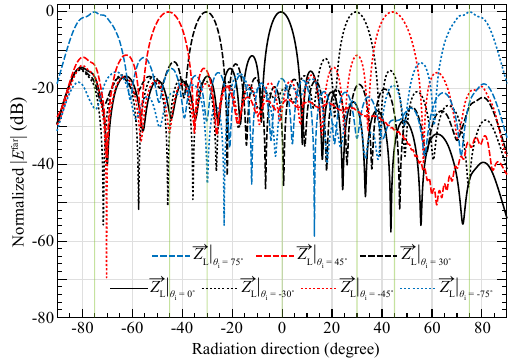}
    \caption{Radiation patterns of the same metasurface with different sets of optimized load impedance values. The operating frequency is 10~GHz. The vertical green solid lines represent desired radiation directions. The radiation patterns are normalized to their maximum values respectively. The green vertical lines indicate the designed main beam direction.}
    \label{fig:scanning_method2}
\end{figure}

The first scanning method  requiring a change in working frequency typically provides narrow scanning angle ranges. When the working frequency deviates significantly from the designed frequency, the performance of the metasurface will deteriorate, especially when the expected radiation direction is very oblique. To handle this problem, the scanning method II can be used. Here, we tune the impedances of reconfigurable loads between several precomputed sets of values while the operating frequency is fixed. 
The radiation patterns for the same metasurface under this scanning method are shown in~\figref{fig:scanning_method2}. In this example, the radiation direction is tuned from $-75\degree$ to $75\degree$. Generally, this scanning method can cover the full angular range over the ground plane.

To fully leverage the attributes of both scanning methods, Scanning Methods I and II can be jointly utilized. Scanning Method II provides ``coarse'' scanning across a broad spatial angle, making it useful for object detection, while Scanning Method I offers a faster scanning technique within a narrower spatial angle, ideally suited for object tracking. Theoretically, this combined approach can result in highly efficient  beam scanning.

\section{Modular approach for multifunctional metasurfaces}
\label{sec:section6}
We have shown that a subwavelength strip array can effectively convert a propagating wave into surface waves. Next, we investigate how to reconfigure the array for guiding surface waves. In this study, we assume that the total number of strips is equal to $2 N$. The first half of the strips are denoted as the first section and the remaining $N$ strips are denoted as the second section. Only the first section of the entire strip array is illuminated by propagating waves. The first section works as a surface wave launcher, while the second section works as a receiver. Here, we do not use a plane wave as an incident propagating wave anymore, as the electric field at the edge of the first section of the array significantly affects the power received by the second section. 
We assume that the first section is illuminated by a Gaussian beam which should be converted into surface waves. 
By properly controlling the waist radius of the Gaussian beam, the field at the end edge of the first section of the strip array will be close to zero, and the effect of the reflection from the bare ground plane can be neglected. Here, for simplicity, the incident wave propagates along the $z$-axis and is specified as~\cite{Comsol2023wave}
\begin{equation}
    \mathbf{E}_{\rm inc}^{\rm g} (y,z) = E_0 \sqrt{\frac{w_0}{w(z)}} e^ {-\frac{y^2}{w^2(z)} -j k_0 z -j k_0 \frac{y^2}{2 R(z)} +j \frac{\eta(z)}{2} }  \hat{x},
\end{equation}
where $E_0=1$~V/m, $w(z)=w_0 \sqrt{ 1+\left( \frac{z-p_0}{z_0} \right)^2 }$ represents the spot radius for different positions along the propagation axis, the focal plane is placed on the PEC ground plane, i.e., at $p_0=0$, and the Rayleigh range $z_0 = \frac{k_0 w_0^2}{2}$, while the Gouy phase shift is given by $\eta(z)=\arctan \frac{z-p_0}{z_0} $. The minimum input beam waist radius is set as $ w_0 = Nd/6=13\lambda_0/12$. Here, it is assumed that the delayed fields at the edge of the axis away from the first half of the strip array, where the field amplitude falls to 0.012\% of the peak value, are essentially zero. 

Before delving into guiding the received surface wave, let us first assume that the strip array only has the first section ($N$ strips in total, i.e., 52 strips). To simplify the simulation model, here the center axis of the strip array is shifted to coincide with the $z$-axis, while the Gaussian beam is along the $z$-axis, i.e., at normal incidence. The design process for conversion of a Gaussian beam is similar to the above-considered plane-wave illumination. Efficient conversion of a Gaussian beam to surface waves is also achieved by globally optimizing the load impedance distribution with the same objective function shown in~\eqref{eq:objective_function}. 
Equation~(\ref{eq:matrix_Ohm}) is used to obtain the induced currents. 

\begin{figure}[!ht]
    \centering
    \includegraphics{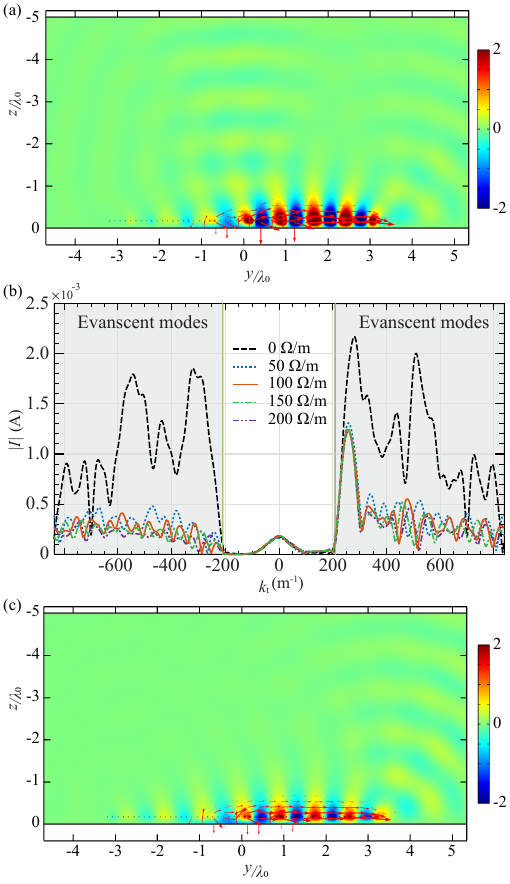}
    \caption{
    Real part of the scattered electric field (with units of V/m) for lossless receiving array (a) and for an array with the load impedances having the real parts $100~{\rm \Omega/m}$ (c). Panel (b) shows the induced current amplitude as a function of the tangential component of the wavevector for load impedances with different values of resistance per unit length.
    }
    \label{fig:full_illumination_gaussian}
\end{figure}

For the case where all the strips except the last one are connected with pure reactive load impedances, the real part of the scattered field is shown in~\figref{fig:full_illumination_gaussian}(a).
The required load impedance distribution (except the one in the last strip) for such conversion is shown in~\figref{fig:surface_wave_guide_gaussian}(a)  with the  shaded area, where the last strip is loaded by a complex impedance per unit length $2085.6-j40280.7~{\rm \Omega/m}$. The conversion efficiency calculated by~\eqref{eq:conversion_efficiency} gives 94.6\%~(Matlab), while the simulation results of COMSOL Multiphysics give 94.9\%. 
We see that in this case the whole array is resonant, as there is a superposition of forward- and backward-traveling surface waves; see the black dashed curve in~\figref{fig:full_illumination_gaussian}(b). 

As has been shown above, the resonance can be suppressed by introducing loss into the load impedance densities. We also have a similar observation for Gaussian beam illumination, as depicted in~\figref{fig:full_illumination_gaussian}(b). Here, we chose the real parts of all load impedance densities (except the last one) equal 100~${\rm \Omega/m}$.
The optimized imaginary parts of the load impedances  are depicted in the shaded area in~\figref{fig:surface_wave_guide_gaussian}(c).
In this case,  the optimized load impedance of the last strip is  $6409.5-j43661.0~{\rm \Omega/m}$.  
 The corresponding scattered electric field distribution is depicted in~\figref{fig:full_illumination_gaussian}(c).  The efficiency found from 
 ~\eqref{eq:conversion_efficiency} gives 82.5\%~(Matlab), while the simulation results of COMSOL Multiphysics give 85.2\%.

\section{Guiding the surface wave and reradiating it as propagating waves}
\label{sec:section7}
In practical applications, the surface wave often needs to be guided. To show a possibility to reconfigure arrays for this funcion, we consider strip arrays formed by  two cascaded sections, where each section has $N$ strips. Specifically, the first section serves as a receiving leaky-wave antenna and a surface wave launcher, while the second section operates as a surface-wave waveguide and/or propagating-wave launcher. Here, for simplicity, the central axis of the first section of the strip array coincides with the $z$-axis. The first section is illuminated by a Gaussian beam. The required load impedances for conversion from the Gaussian beam to surface waves are already known, as mentioned in the previous section. The required load impedances of the second section are found by a global optimization.

\subsection{Surface-wave waveguide}

When the second section of the strip array works as a surface-wave waveguide, it should collect as much power as possible and carry it from left to right. Hence, the objective function is also given by equation~\eqref{eq:objective_function}. Now we have $2N$ strips and only the load impedances of the second section need to be optimized, as the first section is already known. This is totally different from~\cite{budhu2024near}, where the optimization is done for all unit cells.

\begin{figure}[!ht]
    \centering
    \includegraphics{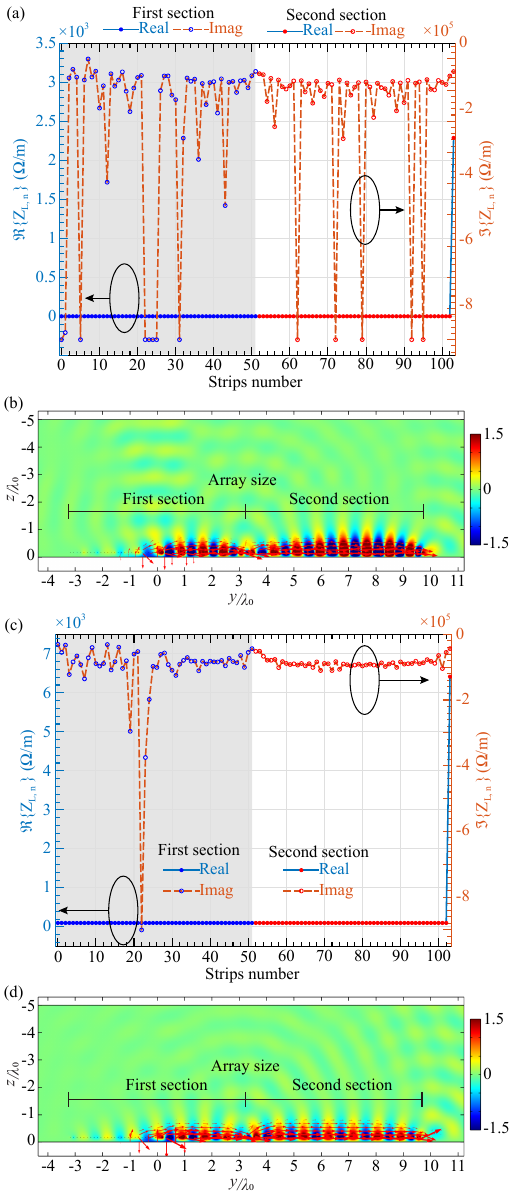}
    \caption{(a) and (b): The load impedances and the scattered electric field distribution (with units of V/m) for lossless surface waveguide. (c) and (d): the same for the real parts of the load impedances equal $100~{\rm \Omega/m}$.}
    \label{fig:surface_wave_guide_gaussian}
\end{figure}

First, we consider the scenario where all the strips except the last one are loaded with lossless impedances. The first  section is the same as in~\figref{fig:full_illumination_gaussian}(a) with the only difference that we set the real part of the load impedance of the $N$-th strip to zero.
By optimizing the load impedances of the second section (only the last $2N$-th strip is loaded with a complex load impedance), we find the required load impedance distribution depicted in~\figref{fig:surface_wave_guide_gaussian}(a), while the scattered electric field is depicted in~\figref{fig:surface_wave_guide_gaussian}(b). The total conversion efficiency is as high as 93.1\% (COMSOL), while the Matlab model gives 91.9\%. We observe that this high total conversion efficiency corresponds to a resonant regime with a superposition of forward- and backward-traveling surface waves. 

To dampen the resonance, lossy load impedances can be introduced. As is shown in~\figref{fig:full_illumination_gaussian}(b), the backward-traveling surface waves are effectively suppressed by adding resistance per unit length equal to 100~${\rm \Omega}$/m. We add the real parts equal to 100~$\rm \Omega$/m to all previously lossless strips and optimize the load impedances of the second section of the strip array. In other words, the variables to be optimized are the imaginary parts of the load impedances of the whole second section and the real part of the load impedance of the last ($2N$-th) strip. The determined load impedance distribution is depicted in~\figref{fig:surface_wave_guide_gaussian}(c), while the real part of the scattered electric field is shown in~\figref{fig:surface_wave_guide_gaussian}(d). The conversion efficiency is 70.4\% (COMSOL model) and  68.4\% (Matlab model). The fluctuations of the required load impedance of the second section in~\figref{fig:surface_wave_guide_gaussian}(c) are much smaller compared to those in in~\figref{fig:surface_wave_guide_gaussian}(a), and the surface wave moves smoothly to the right direction. 

\subsection{Anomalous reflectors}
Apart from the function of guiding surface waves, the second section can also work as 
a launcher of a propagating wave traveling in a desired direction. This way the whole device realizes the functionality of an anomalous reflector, where a wave incident on the first section is effectively reflected to an arbitrary direction when launched from the second section~\cite{achouri_spacewave_2018a}. Regarding the function of anomalous reflection, the objective function is defined as
\begin{equation}
    \mathcal{O}=- {\rm max.} \left\{ \left| E^{\rm far}(\theta_{\rm r},\vec{I}(\vec{Z}_{\rm L})) \right| \right\},
\end{equation}
where $\theta_{\rm r}$ represents the desired radiation direction.

\begin{figure}[!ht]
    \centering
    \includegraphics{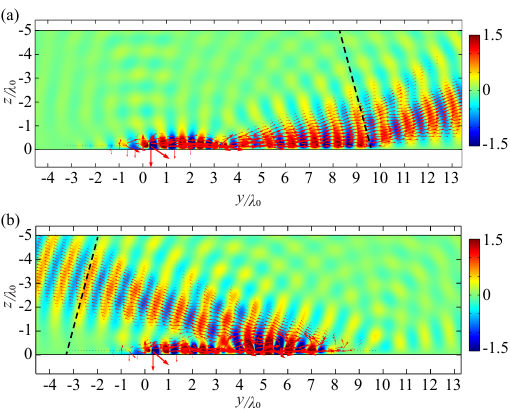}
    \caption{The real part of the scattered electric field $\Re\{E^{\rm sca}_x \}$~[V/m] distributions over space for the desired radiation angle equal (a) $75\degree$ and (b) $-75\degree$.}
\label{fig:anomalous_reflection}
\end{figure}

The scattered electric field of two examples that correspond to the desired radiation angles equal $75\degree$ (forward radiation) and $-75\degree$ (backward radiation) are shown in subfigures~\ref{fig:anomalous_reflection}(a)  and (b), respectively. Here, the impedance loads of the first section of strips correspond to the result in~\figref{fig:full_illumination_gaussian}(a) after removing the real part of the $N$-th strip load impedance. The variables to be optimized are the load impedances of the second section, in which only the load impedance of the last strip has a non-zero real part. 
The efficiency of the anomalous reflection is evaluated by integrating the Poynting vector of the scattered field along the dashed line shown in~Figs.~\ref{fig:anomalous_reflection}(a) and (b),  and then dividing it by the input power. 
The anomalous reflection efficiency for  $75\degree$ (forward radiation) equals 94.6\%, while for $-75\degree$ (backward radiation) 87.9\%. 

\begin{figure}[!ht]
    \centering
    \includegraphics{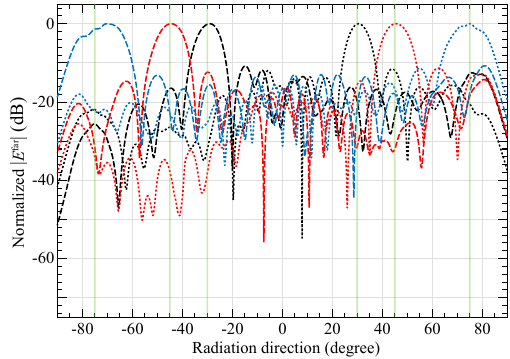}
    \caption{Normalized far-field radiation patterns for different radiation directions. The green vertical lines indicate the desired radiation directions.}
    \label{fig:far_field_pattern_without_lossy}
\end{figure}

The normalized radiation patterns for the case of lossless arrays for different radiation directions are depicted in~\figref{fig:far_field_pattern_without_lossy}. By changing the load impedances, the beam can be scanned over the \textit{full} upper space. 

\subsection{Focusing lens}
\begin{figure}[!ht]
    \centering
    \includegraphics{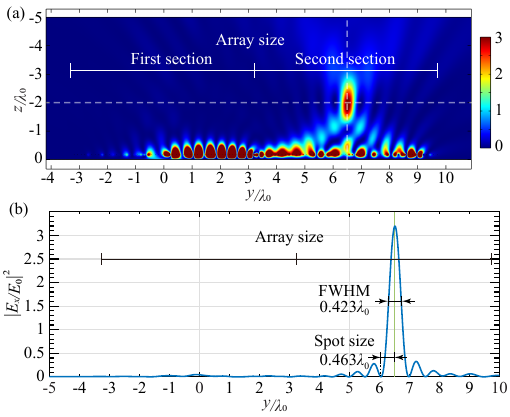}
    \caption{Intensity of the scattered electric field distribution $\left| E_x^{\rm sca} (\hat{y}_0,\hat{z}_0) \right|$~[V/m]$^2$ (a) over the  space and (b) along the focal plane (see the horizontal white dashed line in subfigure~(a)). }
    \label{fig:focusing}
\end{figure}

The second section of the array can act also as a focusing lens for the scattered field. For this  functionality, the objective function is defined as
\begin{equation}
    \mathcal{O} = -{\rm max.} \left\{\left| E_x^{\rm sca} (\hat{y}_0,\hat{z}_0) \right| \right\},
\end{equation}
where the coordinate $(\hat{y}_0,\hat{z}_0)$ represents the position of the desired focal point. Here, the strips of the first section are loaded by impedances that correspond to the result in~\figref{fig:full_illumination_gaussian}(a) shown in the shaded area. The variables to be optimized are the load impedances of the second section, in which only the load impedance of the last strip has a non-zero real part.  To show a near-field focusing possibility, the focal point is chosen at $(6.5\lambda_0, -2\lambda_0)$. The intensity of the scattered electric field is shown in~\figref{fig:focusing}(a). The position of the focal point is placed at the central axis (see the vertical white dashed line in~\figref{fig:focusing}(a)) of the second section of the strip array, and the focal plane (see the horizontal white dashed line in~\figref{fig:focusing}(a)) is away from the PEC ground plane by $2 \lambda_0$. It is important to note that the position of the focal point can be controlled by tuning the  load impedances.

The reflection efficiency is calculated by integrating the Poynting vector along the horizontal white dashed line as is shown 
in~\figref{fig:focusing}(a) and dividing by the input power, which gives 96.9\%. The spot size is measured from the central maximum to the first minima in the diffraction pattern, which gives $0.463\lambda_0$.  The FWHM gives $0.423\lambda_0$. The focusing efficiency is defined  as the flux of the Poynting vector over the spot size along the focal plane normalized by the input power, which gives 83.1\%.

\section{Conclusion}
\label{sec:section8}
To conclude, an effective method for converting propagating waves into surface waves has been demonstrated, which can be used for designing propagating/surface wave converters and bi-directional scanning LWAs. The superdirective conversion is realized by using subwavelength strip array, and the required load impedances for various conversions are found by globally optimizing the load impedances. The demonstrated high and superdirective conversion efficiency is often accompanied by a  resonance of the whole array. This resonance can be suppressed  by introducing loss into load impedances,  if the goal is to excite a single surface wave traveling in one direction. 

The designed LWAs allow two different scanning methods. One of them is realized by tuning the operational frequency, and the other by tuning the load impedances of the unit cells. From the theoretical point of view, the considered arrays have the ability to scan into all directions on the broadside of the strip array.  
The investigation on guiding surface waves is made for illuminations by a Gaussian beam, by cascading two sections of strips array with the same structure, that is, by fixing the load impedances of the first section and globally optimizing the second section. The first section works as a surface wave launcher, while the second section can play the role of a surface wave waveguide, an anomalous reflector, or a focusing lens. 
Our findings promise to be a valuable asset for many 
applications in the field of antenna and microwave engineering.




\section*{Acknowledgments}
The authors thank Dr. Nemanja Stefan Perovi{\'c}, Prof. Xuchen Wang, Dr. Francisco Cuesta, and Dr. Xiaojie Zhu for helpful discussions. 
V.A. acknowledges the Finnish Foundation for Technology Promotion. V.A. and S.T. acknowledge Research Council of Finland within the RCF-DoD Future Information Architecture for IoT initiative under grant no. 365679.


%
\bibliographystyle{IEEEtran}
\bibliography{IEEEabrv,m_refs}

\newpage

\end{document}